\documentclass[11pt,a4paper]{article}
 \usepackage[dvips]{graphics}
 \usepackage{graphicx}
 \usepackage{afterpage}
 \usepackage{enumerate}
  \usepackage{longtable}

\begin{document}
\small
 \title{\bf Effect of the anomalous dispersion in the solar atmosphere on results of magnetic field measurements by the line-ratio method
}
\author{\bf V.G. Lozitskii$^1$ and V.A. Sheminova$^2$ }
  \date{}

 \maketitle
 \thanks{}
\begin{center}
{ $^1$Astronomical Observatory, Shevchenko National University,\\ Observatornaya 3, Kyiv, 04053, Ukraine\\

      $^2$Main Astronomical Observatory, National Academy of Sciences of Ukraine,\\ Akademika  Zabolotnoho 27,  Kyiv,  03143, Ukraine,  e-mail: shem@mao.kiev.ua
}
\end{center}

\begin{abstract}
{On the basis of Stokes parameter calculations for the Fe I $\lambda\lambda$ 524.7 and 525.0 nm lines and the Holweger-Muller model atmosphere, the effect of the anomalous dispersion on solar magnetic field measurements by the line-ratio method is analyzed. It is shown that with the present-day observational accuracy the anomalous dispersion should be taken into consideration in the line-ratio method only when the following four conditions are fulfilled simultaneously: a) the inclination of the magnetic lines to the line of sight does not exceed 20$^\circ$; b) the magnetic field strength is larger than 100 mT; c) the cross profile of the magnetic field in subtelescopic flux tubes is rectangular; and d) the parts of the magnetically sensitive line profiles close to the line center ($\Delta\lambda \leq 4$ pm) are used.

}
\end{abstract}


\section{Introduction }

The line-ratio method [8] is used for measuring magnetic fields in subtelescopic ($d \leq 10^2$ km) structures on the Sun. The basic idea consists in comparison of magnetograph signals for spectral lines that have the same depth of formation and temperature sensitivity, but have different Lande factors. When regions with low field strengths ($H \leq 30$--50 mT) fall on the entrance slit, such lines give the same values of $H_\parallel$ for the measured longitudinal fields. But if there are areas with strong ($H > 50$--100~mT) fields in the region, the $H_\parallel$ found with different lines will differ, as the relationship between the magnetograph signal and the actual field strength is nonlinear.

The method does not depend on the spatial resolution in direct observations, but it requires precise information about the thermodynamic characteristics of the medium where the subtelescopic magnetic structures are localized. On account of this, the total number of free parameters (both magnetic and non-magnetic) is approximately 10, and this makes some simplifying assumptions necessary. The most frequently used assumptions are: the magnetic field is longitudinal, the radial velocities are the same inside the small-scale flux tubes and outside them, and the contribution from the anomalous dispersion (AD) is insignificant. The last assumption has not been substantiated by rigorous quantitative calculations with reference to the theory and practice of the line-ratio method. The purpose of this study is therefore to find out whether it is permissible (and under what conditions) to neglect the effect of AD when small-scale fields are measured.

\section{Calculation of the theoretical Stokes profiles}

Formation of absorption lines in the presence of a magnetic field is described with the transfer equations for polarized light. The polarized radiation is usually given in a parametric representation. According to the theory of Stokes, who was the first to introduce the parametric representation of the polarized radiation, the intensity and polarization state are defined by four Stokes parameters, $I$, $Q$, $U$, and $V$. The physical meaning of these becomes clear from the equations:
\begin{eqnarray}
I &  = & I_0+I_p=I_0 + \sqrt{Q^2+U^2+V^2},\nonumber\\
 Q &  =&I_{lin}(\varphi=0^{\circ})-I_{lin}(\varphi=90^{\circ}),\nonumber \\
 U &  =&I_{lin}(\varphi=45^{\circ})-I_{lin}(\varphi=135^{\circ}), \\
 V &  =&I_{circ, right}-I_{circ, left}\nonumber
\end{eqnarray}
\noindent where $ I_0,~I_p$ are the intensities of the unpolarized and polarized components, respectively, $I_{lin}$ and $I_{circ}$ are the intensities of the linearly and circularly polarized components; $\varphi$ is an angle reckoned from the $OX$ direction in the $XOY$ plane perpendicular to the line if sight. The choice of coordinate system and the magnetic field vector orientation affect the Stokes parameters. The orientation of the magnetic field is often chosen in the way proposed by Shurcliff [11]: an arbitrary direction of the vector \textbf{H} is determined by the inclination angle $\gamma$, which is reckoned from the $OZ$ axis to the direction of \textbf{H}, and the azimuth $\varphi$ which is reckoned from the $OX$ axis to the direction of the vector \textbf{H} projection on the plane $XOY$.

The theory of absorption line formation in the magnetic field was first developed by Unno [16] and later was generalized by many authors: Stepanov, Rachkovskii, Obridko, Stenflo, Staude, Domke, Landi Degl'Innocenti, et al. By now, it has been developed in such a detail that it can provide quite a reliable basis for the theoretical interpretation of magnetographic and polarimetric observations. Analytical methods for solving the transfer equations developed rapidly in the late 1960s and in the 1970s, but their application is always based on approximations that restrict the class of astrophysical problems. For example, it is impossible for realistic model atmospheres to indicate where magnetically sensitive lines are formed, what magnetic field and velocity gradients exist, etc.

Progress in high-precision observations of four Stokes parameters with Stokes polarimeters [15] has increased the need for numerical solutions of transfer equations, as well as for the mathematical software for theoretical calculations of the Stokes parameter profiles. In this field of research, the works by Mattig, Beckers, Wittmann, Staude, Landi Degl'Innocenti, van Ballegooijen, et al., are well known. On this basis, Sheminova wrote the SPANSATM program [9,10], which includes all achievements of other authors, offers the necessary service facilities, and is very helpful. The only substantial limitation is the assumption of the local thermodynamic equilibrium (LTE). Non-LTE effects may nevertheless be taken into account empirically, through the coefficients of deviation from LTE.

Magneto-optical effects in the theory of spectral-line formation in the presence of a magnetic field were taken into account for the first time by Rachkovskii [5,6]. When the transfer equations are solved by numerical methods, accounting for these effects does not cause additional difficulties, as the anomalous dispersion coefficients appear in the transfer equations through the anti-symmetric elements of the absorption matrix. Below we shall give the principal formulas in order to follow the contribution of the anomalous dispersion to the transfer equations. These equations for polarized radiation with the AD allowed for have the following vectorial form:
\begin{equation}
 \frac{d\bf{I}}{d\tau} = \frac{1}{\mu}[(\bf{\eta_0}
+ \bf{\eta)}\bf{I} - (\bf{\eta_0 B}+\bf{\eta S})]
\end{equation}
\noindent where
\begin{equation}
 \bf{I}=\left(\begin{array}{c}
I\\ Q\\ U\\ V
\end{array}  \right),~~~
 \bf{S}=\left(\begin{array}{c} S\\ S\\ S\\ S
\end{array}  \right),~~~
 \bf{B}=\left(\begin{array}{c} B\\ 0\\ 0\\ 0
\end{array}  \right),
\end{equation}
\begin{equation}
 \bf{\eta_0}=\left(\begin{array}{cccc}
\eta_0& 0& 0& 0\\ 0& \eta_0& 0& 0\\ 0& 0& \eta_0& 0\\ 0& 0& 0&
\eta_0
\end{array}  \right),~~~~~~
\bf{\eta}=\left( \begin{array}{cccc} \eta_I&  \eta_Q& \eta_U&
\eta_V\\ \eta_Q& \eta_I& \rho_V&-\rho_U\\ \eta_U& -\rho_V&\eta_I&
\rho_Q\\ \eta_V& \rho_U& -\rho_Q& \eta_I
\end{array}  \right).
\end{equation}
Here $\mu= \cos\theta$; $B$ is the Planck function; $S$ is the source function in a line; $\eta_0$, is the ratio of the selective absorption coefficient at the line center $k_{\lambda_0}$ to the continuous absorption coefficient $\kappa_5$ at the wavelength $\lambda = 500$ nm; $\eta_I,~ \eta_Q,~ \eta_U,~\eta_V$ are the ratios of the selective absorption coefficients for four Stokes parameters to the coefficient $\kappa_5$, and $\rho_Q,~ \rho_U,~ \rho_V$ are the anomalous dispersion coefficients as ratios to the parameter
 $\kappa_5$. The quantities  $\eta_I,~ \eta_Q,~ \eta_U,~\eta_V$ and $\rho_Q,~ \rho_U,~ \rho_V$ are determined by the direction of the magnetic lines of force, i.e., by the inclination $\gamma$, azimuth $\varphi$, and also by the coefficients of selective absorption and anomalous dispersion for radiation linearly polarized in the direction of the field ($\eta_p,~ \rho_p$), counterclockwise polarized in the plane perpendicular to the field direction ($\eta_b,~\rho_b$), and, finally, clockwise polarized in the same plane ($\eta_r,~\rho_r$). According to [17], the latter coefficients are defined by:
\begin{eqnarray}
\eta_p&= &\frac{k_{\lambda_0}}{\kappa_5}[ H(a,
v) + \frac{1}{2}\frac{1}{\Delta\lambda_D^2}\frac{\partial^2H}{\partial v^2}v_0],
\nonumber\\
\eta_{r,b}&=&\frac{k_{\lambda_0}}{\kappa_5}
 [H(a,v \mp v_H) + \frac{1}{2}\frac{1}{\Delta\lambda_D^2}\frac{\partial^2H}{\partial v^2}v_1]
\nonumber\\
\rho_p&=&\frac{k_{\lambda_0}}{\kappa_5}
[F(a,v) + \frac{1}{2}\frac{1}{\Delta\lambda_D^2}\frac{\partial^2F}{\partial v^2}v_0],
\nonumber\\
\rho_{r,b}&=&\frac{k_{\lambda_0}}{\kappa_5} [F(a,v \mp
v_H) + \frac{1}{2}\frac{1}{\Delta\lambda_D^2}\frac{\partial^2F}{\partial v^2}v_1]
\end{eqnarray}
where $H(a, v)$ and $F(a, v)$ are the Voigt and Faraday functions, respectively; $v$ is the distance from the line center; $v_H$ is the Zeeman shift; $v_0$ and $v_1$, are the distances of the displaced $p$-components and $b$- and $r$-components, respectively. The Faraday function, which is also called the dispersion function, and the Voigt function are defined by
\begin{equation}
H(a,v)=\frac{a}{\pi}\int\limits_{-\infty}^{\infty}\frac{\exp(-y^2)}{(v-y)^2+
a^2}dy,
\end{equation}
\begin{equation}
F(a,v)=\frac{1}{2\pi}\int\limits_{-\infty}^{\infty}\frac{\exp(-y^2)(v-y)}{(v-y)^2+
a^2}dy .
\end{equation}
Here
\begin{equation}
v=(\lambda-\lambda_0)/ \Delta\lambda_D ,
\end{equation}\nonumber
\begin{equation}
\Delta\lambda_D=\frac{\lambda_0}{c}\sqrt{2RT/m_i+v^2_{\rm micro}} ,
\end{equation}\nonumber
\begin{equation}
a= \Gamma \lambda_0^2/(4\pi c\Delta\lambda_D).
\end{equation}\nonumber
The Faraday function takes the anomalous dispersion in absorption lines into account. Its characteristics are described in [18].

Thus, (2) can be written as a system of four  first order differential equations. Using a fifth order Runge-Kutta-Fehlberg method and the boundary conditions according to [13], we obtained the Stokes parameter profiles for the escaping radiation at the Sun's surface at the center of the disk in units of relative depression (or line depth); they are as follows:
\begin{eqnarray}
 R_I(\Delta\lambda) & = & (I_c-I(\Delta\lambda))/I_c =   1-I(\Delta\lambda)/I_c,\nonumber \\
 R_Q(\Delta\lambda) & =& (Q_c-Q(\Delta\lambda))/I_c =   -Q(\Delta\lambda)/I_c,\nonumber\\
 R_U(\Delta\lambda) & = & (U_c-U(\Delta\lambda))/I_c =   -U(\Delta\lambda)/I_c,\nonumber \\
~~~R_V(\Delta\lambda) & =&  (V_c-V(\Delta\lambda))/I_c =   -V(\Delta\lambda)/I_c
\end{eqnarray}
where $I_c,~Q_c,~U_c, V_c$ are the Stokes parameters for the continuous radiation, which is usually considered to be unpolarized, i.e., $Q_c = U_c = V_c = 0$.

To calculate the Stokes parameters for the concrete lines Fe I $\lambda\lambda$ 524.7 and 525.0 nm (multiplet no 1) that have the effective Lande factors $g_{eff}$ equal to 2.0 and 3.0, respectively, we have used the following input data: the HOLMU model atmosphere [12], microturbulent velocity $v_{\rm micro} = 0.8$~km/s, macroturbulent velocity $v_{\rm macro}= 0$, damping constant $\Gamma = 1.5 \Gamma_{\rm WdW}$, iron abundance $A = 7.64$, oscillator strengths $\log gf$ equal to $-5.03$ and $-4.89$, respectively [1].
 \begin{figure}[h!b]
 \centerline{
\includegraphics [scale=1.0]{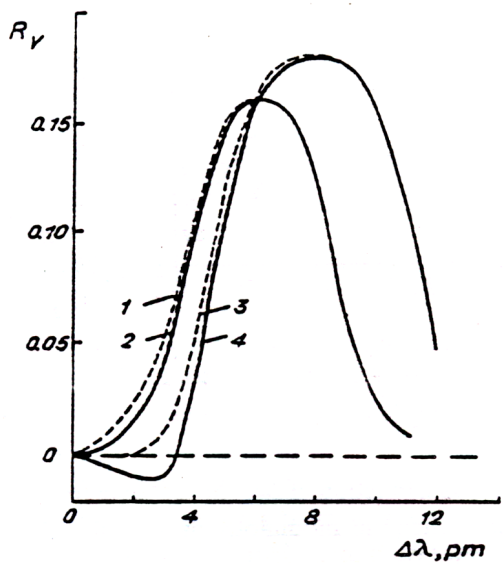}
     }
  \caption {\small
Profiles of the Stokes parameter $R_V$ for the Fe I lines $\lambda$~524.7 nm (curves 1 and 2) and $\lambda$ 525.0 run (curves 3 and 4) with $H$ = 200 mT and $\gamma = 75^\circ$. Profiles 1 and 3 are calculated ignoring the anomalous dispersion, and profiles 2 and 4 take it into account.
}

\end{figure}

Figure 1 shows some calculations with an ES-1061 computer. It is clear that the anomalous dispersion has an appreciable effect only on the parts of line profiles that are close to the line center ($\Delta\lambda\leq 4$--6 pm). As was already noted in [9,10], the wavelength  $\lambda$, excitation potential $EP$, equivalent width $W$, and factor $g_{\rm eff}$ have a minor effect on the anomalous dispersion. Nevertheless, the latter increases with increasing $\lambda$, decreasing $EP$, increasing $W$, and increasing $g_{\rm eff}$. The anomalous dispersion is more sensitive to the parameters of the medium. It increases with decreasing $v_{\rm micro}$, with decreasing $\Gamma$, with the rise of $T$, with increasing $H$, and with increasing $\gamma$. The anomalous dispersion is, in fact, proportional to the magnetic gain of these parameters.

\section{Calculation scheme in the line-ratio method}.

Diagnostic relations in the line-ratio method were calculated according to the scheme described in detail [2]. The profiles $R_V(\Delta\lambda)$ found earlier were used for determination of the ratios
\begin{equation}
r=H_{\parallel}(525.0)/H_{\parallel}(524.7)
\end{equation}
as functions of the distance from the center of the line, $\Delta\lambda$, using the following expressions:
\begin{equation}
H_{\parallel}=H_c\delta_{\parallel}/\delta_c
\end{equation}
where $H_c = 2.14 \cdot 10^7 \Delta\lambda_c/ g_{\rm eff} \lambda^2$ ($H_c$ is in units of T, $\lambda$ and $\Delta\lambda$ are in units of nm), and
\begin{equation}
\delta_\parallel = 2\alpha x^{-2}_m \int_0^{x_m} {R_V(\Delta\lambda, x)xdx},
\end{equation}
\begin{equation}
\delta_c = R_I(\lambda+\Delta\lambda_c)-R_I(\lambda-\Delta\lambda_c).
\end{equation}
Here $R_I(\lambda\pm\Delta\lambda_c)$ is the line profile unperturbed by the magnetic field and shifted by $\pm\Delta\lambda_c$ for calibration of the magnetograph signal $\delta_\parallel$; $\alpha$ is the fraction of the aperture area occupied by flux tubes (filling factor); $x = l/l_0$ is the distance from the axis of symmetry of a flux tube expressed in relative units ($l$ is the line distance and $l_0$ is a typical radius of a flux tube); $x_m$ is the distance at which the field strength becomes zero.

The function $R_V(\Delta\lambda, x)$ under the integral depends on the Zeeman splitting
\begin{equation}
\Delta\lambda_H=4.67\cdot 10^{-8}g_{\rm eff}\lambda^2 H
\end{equation}
where $H$ may, in its turn, also depend on $x$, i.e, $H = H(x)$. Later we shall call the $H(x)$ function a field cross profile (or simply a field profile) in flux tubes.

Expression (14) corresponds to the total magnetic flux is concentrated within flux tubes. But it may be easily generalized to the case when the contribution of the background field of the strength $H_i$ is not zero. Then
\begin{equation}
\delta_\parallel = 2\alpha x_m^{-2} \int_0^{x_m} R_{V,f}(\Delta\lambda, x)xdx+ (1-\alpha)R_{V,i}(\Delta\lambda)
\end{equation}
where indices $f$and $i$ refer to flux tubes and the background field, respectively.

\section{Anomalous dispersion effect on the line-ratio method.}

According to calculations, the assumption that the magnetic field is longitudinal is quite admissible, particularly for $\Delta\lambda < 8$ pm (Fig. 2).  Experimental values $r$ are determined with an error of approximately 5--10\%, and, therefore, a small actual discrepancy between thi relations, for example, for $\gamma = 0^\circ$ and $\gamma = 75^\circ$ at $\Delta\lambda < 8$ pm has no practical importance.

 \begin{figure}[h!]
 \centerline{
\includegraphics [scale=1.0]{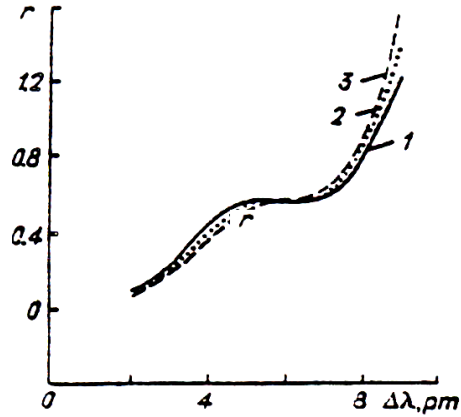}
     }
  \caption {\small
Ratio $r=H_{\parallel}(525.0)/H_{\parallel}(524.7)$ as a function of $\Delta\lambda$, the distance from the line center, for a uniform magnetic field of $H = 200$~mT when the anomalous dispersion is absent and the values of the angle  $\gamma$ are $0^\circ$ (1), $45^\circ$ (2), and $75^\circ$  (3).
}
\end{figure}
 \begin{figure}[h!b]
 \centerline{
\includegraphics [scale=1.0]{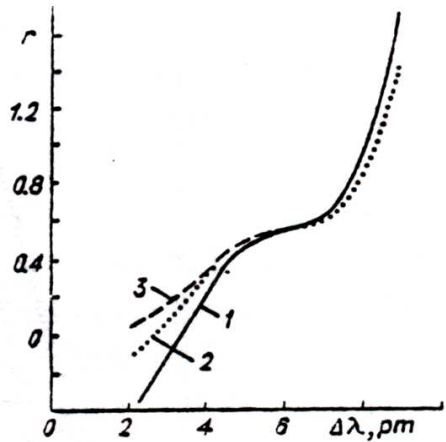}
     }
  \caption {\small
The same as in Fig. 2, but taking the anomalous dispersion into account and for $\gamma = 75^\circ$ (1), $45^\circ$ (2), and $0^\circ$ (3).
}

\end{figure}

 \begin{figure}[h!b]
 \centerline{
\includegraphics [scale=1.0]{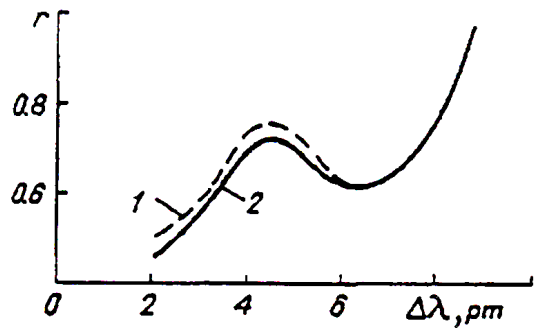}
     }
  \caption {\small
 The same as in Figs. 2 and 3, but for the two-component model $H(x)$ = const = 200 mT, $H_i = 20$ mT, $\alpha= 0.2$, $\gamma = 75^\circ$: ignoring the anomalous dispersion (1), taking it into account (2).
}

\end{figure}

In general, the anomalous dispersion affects $r$ more strongly than the field inclination angle, especially near the line centers ($\Delta\lambda < 4$ pm), as well as at large angles ($\gamma > 20-30^\circ$), in an intense field ($H > 100$ mT), and when the intensity is the same in all points in the aperture (Fig. 3). The contribution of anomalous dispersion is to be taken into account only when the conditions indicated above are fulfilled simultaneously. If at least one of the conditions is not fulfilled, the anomalous dispersion effects will be within the typical errors of determination of $r$.

In particular, if regions with both strong ($H > 100$ mT) non-longitudinal field and non-longitudinal field of small and medium strength ($H \leq 30$--50 mT) fall on the entrance slit, $r$ will be the same within the error limit: for both cases -- when the anomalous dispersion is present or absent -- in the whole interval of actual $\Delta\lambda$. Figure 4 illustrates this, showing the relationships for the two-component model that involves a background field of $H_i = 20$ mT and subtelescopic flux tube with a rectangular field distribution
\begin{equation}
  H(x)={\rm const}= 200~ {\rm mT}.
\end{equation}
Here we adopted, according to the models of [4,7], $\alpha =0.02$, $H_i/\alpha = 100$ mT. The difference of relationships shown in Fig. 4 gives a certain maximum effect for models with a background field because in real flux tubes $H(x) \neq$~const [4].
But if one uses the given method, as in [17], for the study of spatially resolved structures (e.g., pores), when a region with an intense field fills in the aperture area completely, then the neglect of the anomalous dispersion can cause considerable errors, more than 30\% of the true field intensity, for $\Delta\lambda < 4$~pm.

Calculations have shown also that, when the ratios of linear polarization amplitudes, $\delta_\perp = \sqrt{R_Q^2 + R^2_U}$, are used in this method, a situation occurs very similar to the case of circular polarization $R_V$ analyzed above. In particular, the effect of the anomalous dispersion here is also the largest for $\Delta\lambda= 0$--6 pm, whereas it is negligible outside this interval.

\section{Conclusions}

Calculations show that when the line-ratio method is used, the effect of anomalous dispersion cannot be neglected if the following conditions are satisfied simultaneously: a) the inclination angle of the lines of force exceeds $20^\circ$, b) the magnetic field exceeds 100 mT, c) the magnetic field profile in flux tubes is rectangular, and d) parts of line profiles near the center ($\Delta\lambda < 4$~pm) are used. It means practically that when real small-scale flux tubes outside of spots or pores are studied, it is not necessary to take the anomalous dispersion into account not only in the central zone of the solar disk, but even for heliocentric angles of approximately 60--$70^\circ$. Although flux tubes in the solar atmosphere appear to be almost vertical [3,15], the magnetic field profile is nonrectangular in them (according to the data of [4], $H(x) \propto 1 - x^4$). Besides that, a background field is very probable between tubes, and this field transfers a magnetic flux comparable with the flux in the tubes. The criterion stated above is thus violated. This means, with reference to specific proposed models, that neither the model by Rachkovskii and Tsap [17] nor that by Lozitskii and Tsap [4] need be revised from this point of view.

The anomalous dispersion should nevertheless be taken into account in the method when solar pores observed if their size is not smaller than the effective size of the aperture and the heliocentric angles are greater than 20--$30^\circ$.

\end{document}